\documentclass[twocolumn,showpacs,preprintnumbers,amsmath,amssymb]{revtex4}
\usepackage{graphicx}
\usepackage{dcolumn}
\usepackage{bm}
\usepackage{amssymb}
\usepackage{epsfig}
\usepackage{hyperref}
\usepackage{float}
\usepackage[utf8]{inputenc}
\makeatletter
\newcommand*{\rom}[1]{\expandafter\@slowromancap\romannumeral #1@}
\makeatother
\begin{document}
\title{Baryogenesis from B meson  oscillations }
\author{Ann E. Nelson$^{1}$}
\email{anelson@phys.washington.edu}
\author{Huangyu Xiao$^{1}$}
\email{huangyu@uw.edu}
\affiliation{$^1$Department of Physics, University of Washington,  Seattle, WA 98195-1560,USA}

\begin{abstract}

  We show how CP violating B meson oscillations in conjunction with baryon number violating decays can generate the cosmological asymmetry between matter and anti-matter, and   explore the parameter space of a simple,   self-contained model, which can be tested via  exotic B meson decays, and via the  charge asymmetry in semi-leptonic decays of neutral B mesons.
\end{abstract}
\maketitle

\section{Introduction}
{\it Baryogenesis}---generating the cosmological asymmetry between matter and anti-matter--- requires physics beyond the Standard Model (SM). The pioneering work of Sakharov\cite{Sakharov:1967dj} found three  necessary conditions: baryon number violation, C and CP violation, and departure from thermal equilibrium.  Baryon number violation occurs non-perturbatively in the Standard Model\cite{tHooft:1976rip}. CP violation also occurs, however the standard model CP violation appears to be too small to explain the observed baryon asymmetry.  Finally, the  minimal Standard Model contains no mechanism for departure from thermal equilibrium. 

Recent work \cite{McKeen:2015cuz,Ghalsasi:2015mxa} has shown the possibility for low energy baryogenesis via the oscillations of neutral hadrons, in conjunction with new sources of CP and baryon number violation.  In ref. \cite{Ghalsasi:2015mxa}, the oscillating hadrons were {\it mesinos}--bound states of a quark and an anti-squark. In that work a relatively long-lived squark decayed into anti-quarks via baryon number violating R-parity violating decays. A minimal model to capture this physics was studied in detail--that model contained three neutral Majorana fermions ('neutralinos') and a color triplet scalar ('squark').  The same model, in a different parameter region with lighter neutralinos, was shown to lead to baryogenesis via potentially observable baryon and CP violating neutral heavy flavor baryon oscillations\cite{McKeen:2015cuz,Aitken:2017wie}. A similar model, in which baryon number is conserved but also carried by dark matter, was shown to be capable of producing both the  visible matter-anti-matter asymmetry and asymmetric dark matter \cite{Elor:2018twp} via B-meson decays. In the present work, we reexamine the simpler model of ref. \cite{Ghalsasi:2015mxa}, and show that for a different  parameter range that was not considered in the previous work,  baryon number violating decays of $B^0$ mesons are  allowed by experiment, potentially observable, and could be the explanation for baryogenesis. 

The baryogenesis scenario described here begins in the pre-nucleosynthesis early universe with the decays of a long lived scalar into $b-$quarks and anti-quarks. These decays are assumed to take place late enough and at low enough  temperature to allow  hadronization, but before nucleosynthesis.  Most of the $b-$quarks form $B$ mesons. The neutral $B$ mesons then   oscillate and decay, sometimes to baryons or anti-baryons, resulting in the observed asymmetry.

Our model is similar to the one used in ref. \cite{Aitken:2017wie} to produce the baryon asymmetry through oscillations of baryons\cite{McKeen:2015cuz,Ghalsasi:2015mxa}, although even simpler. It is worth noting that even though we use a similar model, the parameter space is  different and we are less constrained  from di-nucleon decay. The mechanism we describe is similar to the one found in   a slightly more elaborate model\cite{Elor:2018twp}, which  could generate a dark matter relic along with   baryon number.   In the current work we have no dark matter sector. We  assume that whatever the dark matter is, it is very weakly coupled  and has no effect on baryogenesis. It would be a straightforward matter to include, for instance,   axion  dark matter. Note that the presence of a late decaying heavy particle can have an effect on the allowed axion  parameter range and substructure\cite{Nelson:2018via}. 

This paper is organized as following: In section \ref{s0} we give an overview of B meson physics and how it could be related to baryogenesis.  A more detailed description of the model and phenomenology is given in section \ref{s1}. In section \ref{s2} we   show that the baryon asymmetry may be produced with  parameters  which are allowed by experiment. In section \ref{s3} we conclude and sketch ideas for future work.

\section{Charge asymmetry in Oscillating B-mesons and sign of the matter-anti-matter asymmetry}\label{s0}
We briefly review the physics of neutral $B-$meson oscillations and describe how new physics in the decays and mixing can lead to baryogenesis.
The oscillations are described by an effective 2 state Hamiltonian:
\begin{equation}\label{Ha}
\mathcal{H}=M-\frac{i}{2}\Gamma=\begin{pmatrix} M&M_{12}\\M_{12}^{*}&M \end{pmatrix}
-\frac{i}{2}\begin{pmatrix} \Gamma &\Gamma_{12}\\ \Gamma_{12}^{*}&\Gamma \end{pmatrix}\ .
\end{equation}
Here $M$ and $\Gamma$ are respectively the dispersive part and the absorptive part of the transition amplitude.
The CP violating phase arg($\Gamma_{12}/M_{12}$) is   reparameterization invariant and  observable in the semi-leptonic charge asymmetry of  neutral meson decays. This phase  is   crucial for our baryogenesis mechanism, as it determines whether there are more $b$ quarks or $b$ anti-quarks at the time of decay. With an asymmetry between $b$ quarks and anti-quarks, the baryon number violating decays of the $b$ quark into two lighter anti-quarks and a Majorana fermion, or the $b$ anti-quark into two quarks and  a Majorana fermion can  produce the observed matter anti-matter asymmetry. In order for baryon number violating decays to produce more matter than anti-matter, there must  be more $b-anti-quarks$ at the time of decay. Because the semi-leptonic decays of $b-anti-quarks$ produce positively charged leptons, this baryogenesis mechanism requires a positive charge asymmetry in either $B^0$  or $B_s$ meson oscillations, or both. The charge asymmetry depends on the phase of the absorptive term ($\Gamma_{12}$) relative to the dispersive term ($M_{12}$) in   B meson mixing.     The   magnitude of $\Gamma_{12}$ and $M_{12}$ are measured    and are in  agreement with the Standard Model in both neutral meson systems. In the Standard Model,    the phase arg($\Gamma_{12}/M_{12}$)  is predicted to be very small  for both the $B^0$  and the $B_s$ due to the unitarity of the CKM matrix and the   relatively small mass of the charm quark, leading to a very small prediction for the charge asymmetry. Experimentally, the charge asymmetry has not yet been distinguished from zero in either system.  The prediction in the standard model is that the charge asymmetry is negative for the $B^0$ and positive but very small for the $B_s$.  Our detailed analysis will show that because more $b$ quarks  fragment into $B^0$ mesons than $B_s$ mesons, and because of the small size of the standard model asymmetry in  $B_s$ mesons, new physics in the the mixing amplitude   may be needed for baryogenesis from $B$ mesons. New physics which makes a    small contribution to $M_{12}$ and is consistent with experimental constraints can    have a   significant effect on the phase arg($\Gamma_{12}/M_{12}$) in either system. When experimental constraints on such new contributions are taken into account, the most promising case for a positive charge asymmetry which is large enough for baryogenesis is a new contribution to mixing in the $B_s$  system.

\section{New Phenomenology from our model}\label{s1} 
A minimal renormalizable model for this baryogenesis mechanism contains three new particles: One Majorana fermion, $\chi$, in the mass range 2-3 GeV,  one new charge -1/3 color triplet scalar $\phi$, with mass in the range 500 GeV to  1.9 TeV,   and a  very weakly coupled  particle, $\Phi$, which decays out of equilibrium into $b$-quarks and anti-quarks.  The upper bound on the $\phi$ mass is from the need for a large enough branching ratio for exotic B meson decay, as we explain later. 
$\Phi$  might be   the inflaton or a string modulus. As   $\Phi$ is very weakly coupled it is not experimentally accessible.  Detailed computation shows that we need the mass of $\Phi$ to be between 11 GeV and about a hundred GeV. We note that it is to be expected that a scalar in this mass range would mainly decay into $b$-quarks. A simple way to achieve this goal is coupling $\Phi$ to the Higgs boson with a very small coupling constant $g$ :$\Delta \mathcal{L}=g\Phi H^{\dagger}H$. The Vacuum Expectation Value (VEV) of the Higgs boson would naturally give rise to $gv \Phi  h$, which results a tiny amount of $\Phi-H$ mixing and a decay channel to b-quarks. 
The neutral Majorana fermion $\chi$ and color triplet $\phi$ will interact with SU(2) singlet  quarks through the following terms:
 \begin{equation}\label{LI}
 \mathcal{L}_{\text{int}}\supset -g_{ij}\phi^*\bar{u}_R^i d^{cj}_R-y_j\phi \bar{\chi}d^{cj}_R+\text{h.c.}
 \end{equation}
 where $i$ and $j$  are flavor indices which run over the three generations of up- and down-type quarks. 
 
The Majorana fermion $\chi$ is neutral and its mass should be greater than the mass difference between proton and electron, $m_p-m_e$= 937.75 MeV, otherwise this model will give rise to proton decay\cite{McKeen:2015cuz}. The mass of $\chi$ will be taken to be 2-3 GeV in this paper.  Mediated by $\phi$, the Majorana fermion $\chi$ could decay to three quarks. The lifetime  must be less than 0.1s to avoid spoiling successful BBN\cite{Jedamzik:2006xz}.  This requires that   $g_{ij} y_{j'}\gtrsim\left(5\text{GeV}/m_{\chi}\right)^5\left(m_{\phi}/350\text{TeV}\right)^2$ \cite{Aitken:2017wie}, with $i=u$ or $c$ and $j,j'= d$ or $s$. 

The colored scalar $\phi$ will either decay to anti-quark pairs or into $\chi$ plus a quark, and could appear in searches for dijet resonances and jets and missing energy. The mass of  $\phi$ should be at least about 500 GeV to pass collider constraints\cite{Aaboud:2017nmi}. There will also be many constraints from nucleon oscillations and di-nucleon decays\cite{Aitken:2017wie}, which give us   upper bounds on various flavor combinations of the $g_{ij}$ and $y_{j}$ couplings.  The strongest bounds, which are on combinations of couplings that allow dinucleon decays \cite{Aitken:2017wie}, requires that $g_{ij} y_{j'}\lesssim \left(m_{\phi}/34\text{TeV}\right)^2$ for $m_{\chi}=2\text{GeV}$, which are consistent with a cosmologically acceptable $\chi $ lifetime if the mass of $\chi$ is greater than $2$GeV. The upper bound on the $\chi$ mass comes from branching ratio of exotic B meson decay, which will be discussed in the following subsection \ref{Bdecay}.
\begin{table}[h!]
\centering
 \begin{tabular}{ |p{0.4cm}||p{0.7cm}|p{1.4cm}|p{1.4cm}|p{1.2cm}|p{1.2cm}|p{1.2cm}|p{1.2cm}|  }
 \hline
 \multicolumn{7}{|c|}{New Particles in Our Model} \\
 \hline
    &Spin&Mass (GeV)&Lifetime&SU(3)&SU(2)&$Q$\\
 \hline
 $\Phi$  & 0    &11-100& 0.2-20 ms  &  Singlet&Singlet& 0\\
 $\phi$ &   0  & 500-1900& &Triplet  &Singlet & -1/3\\
 $ \chi$ &1/2   & 2-3 &$<0.1$ s     & Singlet& Singlet & 0\\
 \hline
\end{tabular}
\caption{This table briefly summarizes the properties of the new particles we introduced. The masses and lifetimes are mainly constrained by colliders and cosmological obeservations, which will be discussed in more detail. The colored scalar $\phi$ will decay rapidly to anti-quark pairs or $\chi$ plus a quark through tree-level couplings.}
\end{table}

Long-lived but unstable particles like $\Phi$ and $\chi$ are extremely hard to detect because the coupling constant is too small, which appears as missing energy in colliders. But the $\chi$ particle is produced in $b$-quark decays and will eventually decay to SM particles, which gives  an opportunity for detection in long lived particle searches\cite{Curtin:2018mvb,Lee:2018pag}.
 
  This model allows B mesons to decay into a baryon  (plus mesons) plus an unstable Majorana fermion. As the latter will equally likely decay into a baryon or an anti baryon  plus mesons, when combined with a CP violating charge asymmetry,  such decays will result in a net baryon number. The box diagrams involving the new particles and the $y_d$ couplings  can also modify the phase arg($\Gamma_{12}/M_{12}$), producing a non-standard charge asymmetry in $B$ oscillations.

\subsection{Exotic B Meson Decays}\label{Bdecay}
\begin{figure}[h] 
 \includegraphics[width=\linewidth]{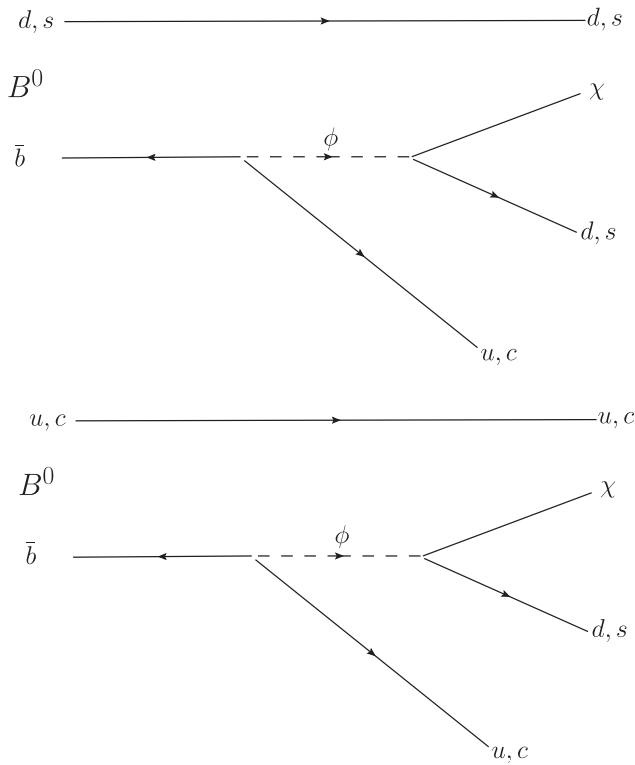}
 \caption{These are the Feynman diagrams for a new $b$ decay channel, which violate baryon number.}
 \label{decay}
 \end{figure}

 There are potentially observable consequences for B meson physics. First, there will be a new decay channel for B mesons, which violates  baryon number, and involves a relatively long lived exotic Majorana fermion. The Feynman diagram for this decay process is shown in Fig.\ref{decay}. 
 
 The Majorana fermion   will be present as missing energy in most collider searches.  Searches for long lived particles, e.g. using a MATHUSLA-like detector, have a chance to find it. \cite{Curtin:2018mvb}. 
 
 In estimating the branching fraction for the exotic decay of $B$ mesons, we assume that the  the $\chi$ mass is low enough that the momenta in the decay are all larger than the QCD scale so that we may treat QCD perturbatively. Then decay may be approximated by  the rate for a heavy $b$ quark  to decay into 3 lighter fermions, while the  light quark in the $B$ meson acts as a spectator.

  The effective interaction term is:
\begin{equation}
\Delta \mathcal{L}=\frac{y_s g_{ub}^*}{m_{\phi}^2}\bar{b}u s\chi .
\end{equation}In this limit, neglecting the masses of the light quarks,  the decay rate is approximately\cite{Aitken:2017wie}:
\begin{equation}
\Delta \Gamma\sim \frac{|y_s g_{ub}^*|^2}{60 (2\pi)^3 m_{\phi}^4}m_b \Delta m^4,
\end{equation}
where $\Delta m$ is the mass splitting between the $\chi$ and the bottom quark.
  The $B^0$,$B^+$, and $B_s$  could all possibly decay to a baryon plus mesons plus  the $\chi$ fermion, with the $\chi$ appearing either as missing transverse momentum or as a long lived particle. As $\chi$ can decay  into either 3 quarks or 3 anti-quarks, it will appear as either a baryon plus mesons or an anti-baryon plus mesons. Note that $y_b$ is not strongly constrained by colliders or by the oscillations of    baryons , therefore this branching fraction can be relatively large.

To produce the required baryon number, we will find in   section \ref{s2} that the branching ratio $\text{Br}_{B\rightarrow\mathcal{B}}$ must be in the range  of $\text{Br}_{B\rightarrow\mathcal{B}}\sim  10^{-3}-10^{-1}$.  From the decay rate we've estimated, the branching ratio is:
\begin{equation}
\text{Br}_{B\rightarrow\mathcal{B}}\sim 10^{-3}\left (\frac{\Delta m}{ 2\text{GeV}}\right)^4 \left(\frac{\text{1TeV}}{m_{\phi}}\frac{\sqrt{y_s g_{ub}}}{0.53} \right)^4.
\end{equation}
   The branching ratio has to be greater than $ 10^{-3}$ to generate sufficient baryon number, which imposes an upper bound  $m_{\phi}\lesssim 1.9 \text{TeV}$ if  we take $\Delta m\sim 2\text{GeV}, y_b g_{us}\sim 1 $. This is compatible with collider constraints on a colored scalar, including the constraints on resonant single $\phi$ production  \cite{Aaboud:2016nwl,Monteux:2016gag,Khachatryan:2014rra}. 

 \subsection{New contributions to  $B^0 - {\bar{B}}^0$ Mixing} 
 In our model, baryogenesis can result from baryon number violation in $B$ meson  decay. The baryon number will depend on CP violating phases determining the charge asymmetry  in $B_d$ and $B_s$ meson mixing. In the SM the CP violating phase in $B^0$-$\overline{B}{}^0$ mixing is determined by the CKM matrix. Our new model could also contributes to $B^0$-$\overline{B}{}^0$. As  $|M_{12}|$ is measured with high precision and agrees well with SM predictions, the new contribution should be much smaller than the standard box diagram. It generically will have a different phase.  Because the CP violating phase is small,  the  width difference $\Delta \Gamma_q$ and $\Delta m_q$ may be used to estimate the magnitude of $\Gamma_{12}$, $ M_{12}$, with $\Delta \Gamma_q=2 |\Gamma_{12}|$ and  $\Delta m_q=2|M_{12}|$, where the subscript $q$ represents arbitrary quarks.  $\Delta m_q$ is well measured\cite{Barberio:2008fa,Abazov:2006dm,Abulencia:2006mq}:
 \begin{equation}
 \Delta m_d=0.5064\pm0.0019\text{ps}^{-1}, \quad \Delta m_s=17.757\pm0.021 \text{ps}^{-1}
 \end{equation}
 The decay width difference $\Delta \Gamma_s$ is given by\cite{Artuso:2015swg}:
 \begin{equation}
 \Delta \Gamma_s=0.085\pm0.015 \text{ps}^{-1}
 \end{equation}
 $\Delta \Gamma_d$ is not well measured because $\Delta \Gamma_d/\Gamma_d$ is too small and the uncertainties are relatively very large. However, in the approximation of negligible CP violation in mixing, the ratio $\Delta\Gamma_q/\Delta m_q$ is equal to the small quantity $|\Gamma_{12}/M_{12}|$, which is independent of CKM matrix elements and could be used to determine $\Delta \Gamma_d$:
 \begin{equation}
 \Delta \Gamma_d=0.0026 \text{ps}^{-1}
 \end{equation}
The CP asymmetry in semileptonic B decays is defined as:
 \begin{equation}
 A_{\text{SL}}=\frac{\Gamma[\bar{B}^0(t)\rightarrow l^+ X]-\Gamma[B^0(t)\rightarrow l^-X]}{\Gamma[\bar{B}^0(t)\rightarrow l^+ X]+\Gamma[B^0(t)\rightarrow l^-X]}
 \end{equation}
 where $X$ stands for any other particles produced in this inclusive process. This asymmetry would be determined by the relative phase between the absorptive and dispersive parts of the transition amplitude (For calculations of transition amplitude in Standard Model, see \cite{Hagelin:1981zk, Buras:1984pq,Cheng:1982hq}), $ A_{\text{SL}}=\text{Im}\Gamma_{12}/M_{12}$.
Assuming only $M_{12}$ receives a new contribution from new physics, the experimental searches for $A_{\text{SL}}$ gives a range of \cite{Aaij:2014nxa,Aaij:2016yze,Klaver:2016sft,Artuso:2015swg}:
 \begin{equation} \label{ASL}
 \begin{split}
 &A_{\text{SL}}^d\in(-5.9\times 10^{-3}, \,-4\times 10^{-4}) ,\\
 &A_{\text{SL}}^s\in (-1.11\times 10^{-3} ,\, 8.8\times 10^{-4}).
 \end{split}
 \end{equation}

 The SM predictions  for  $A_{\text{SL}}$ of both $B^0_s$ and $B_d^0$  are very small\cite{Laplace:2002ik,Artuso:2015swg}: 
  \begin{equation}
 \begin{split}
 &A_{\text{SL}}^d=(-4.7\pm 0.6)\times 10^{-4} \\
 &A_{\text{SL}}^s=(2.22\pm 0.27)\times 10^{-5}
 \end{split}
 \end{equation}
 
 As mentioned, $A_{SL}$ has to be positive to give rise to baryogenesis, which means that we will need a positive asymmetry in the $B_s$ system. We will find that   new physics in the mixing is favored to make this sufficiently positive.
 There is still room for new physics which could make an order of magnitude change in  $A_{\text{SL}}$ in either system.

The second term of the Lagrangian, will directly result in an extra contribution to  $B^0$-$\overline{B}{}^0$ transition amplitude, which can be seen from the box Feynman diagram \ref{box}:
  \begin{figure}[h] 
 \includegraphics[width=\linewidth]{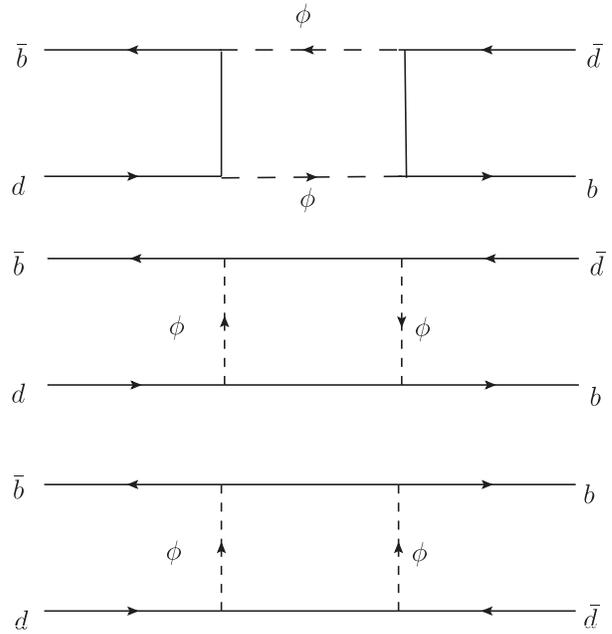}
 \caption{These are the Feynman diagram for a non-standard to B meson oscillations.  }
 \label{box}
 \end{figure}
 
 The dispersive part will be given by the exchange of off-shell particles, while the absorptive part of transition amplitude will be given by the exchange of on-shell particles.  Since $\phi$ is heavy, there will not be any absorptive contribution from the new diagrams as long as the mass of $\chi$ is greater than half of B meson mass, and even if $\chi$ is lighter than this any such contribution would be tiny.   For our purpose, we only need to affect the small phase arg($\Gamma_{12}/M_{12}$), which does not require any new contribution to $\Gamma_{12}$. Any contribution to $M_{12}$ from new physics which has a phase different from the CKM phase could  achieve this. The extra transition amplitude would be given by (see Appendix for detailed calculation):
 \begin{equation}
\Delta M_{12}=0.66\frac{f_B^2m_B B_B }{16\pi^2 m_{\phi}^2}(y_b y_d^*)^2  
 \end{equation}
 where $f_B$ is the decay constant of B mesons, $m_B$ is B meson mass and $B_B$ is a bag parameter, which is order one. The subscript in $y_d^*$  still represents the down type quark.
 
 It is worth noting that the $y_d^i$ are the only coupling constants from the new model that affect transition amplitude. However, the constraints for the parameters from dinucleon decay and heavy flavor baryon oscillations only set a upper bound for combinations of $g_{ud}^{ij}  y_d^k$. From another point of view, the magnitude of transition amplitude is measured with high precision, which agrees with SM predictions roughly and that imposes constraints on $y_d$:
   \begin{equation}
  \begin{split}
 &|y_b y_d^*|^2\lesssim 1.02\times10^{-6}\left(\frac{m_{\phi}}{1\text{TeV}}\right)^2 \\
 &|y_b y_s^*|^2 \lesssim 7.33\times 10^{-7}\left(\frac{m_{\phi}}{1\text{TeV}}\right)^2
  \end{split}
  \end{equation}

  Noting that $(y_b y_s^*)^2 $ is generally a complex number, if it has large phase difference with transition amplitude from SM predictions, this term could give rise to positive value of semileptonic asymmetry. The  B meson transition amplitude  gives the strongest constraint on the combinations $(y_b y_d^*)^2 $ and $(y_b y_s^*)^2 $, therefore we may fit these parameters to the baryon asymmetry.

\section{Cosmological Production of the baryon asymmetry}\label{s2} 
 The massive particles, $\Phi$, which mainly decay out of equilibrium to $b$-$\bar{b}$ quarks, provide the necessary departure from thermal equilibrium  to produce the baryon asymmetry.  The $b$-$\bar{b}$ quarks will quickly hadronize then   decay into other lighter particles. Since the time scale for $b$-$\bar{b}$ quark decay is much shorter than the age of the universe,  in considering the thermal evolution of the universe it is valid to neglect the brief existence of $B$ mesons or baryons and  consider the $\Phi$ decay products to be radiation. The radiation produced by $\Phi$ decays will thermalize  on a timescale that is very short compared with the lifetime of the $\Phi$. Therefore, the evolution of the energy density of radiation and $\Phi$ can be described by the following equations:
 \begin{equation}
\begin{split}
&\frac{d\rho_{\Phi}}{dt}+3H\rho_{\Phi}=-\Gamma_{\Phi}\rho_{\Phi}\\
&\frac{d\rho_{r}}{dt}+4H\rho_{r}=\Gamma_{\Phi}\rho_{\Phi}
\end{split}
\end{equation}
where $\rho_{\Phi}$ is the energy density of $\Phi$ , $\rho_{r}$ is the energy density of radiation, and $\Gamma_{\Phi}$ is the decay rate of $\Phi$.  The energy density of $\rho_{r}$ could directly determine the temperature of the radiation through:
\begin{equation}
\rho_{r}=\frac{\pi^2}{30}g_{*}(T)T^4
\end{equation}
where $g_*(T)$ is the effective number of degrees of freedom. The Hubble parameter is given by:
\begin{equation}
H=\sqrt{\frac{8\pi}{3} \frac{\rho_r+\rho_{\Phi}}{M_{\text{pl}}^2}}
\end{equation}
where $M_{\text{pl}}=1.22\times 10^{19} \text{GeV} $  is the Planck mass.  Solving these equations numerically we can obtain the thermal history of the early universe during the early matter dominated era. 

When the temperature is below  a scale of order $\Lambda_{\text{QCD}}\sim 200$ MeV, the quarks will hadronize. The b quarks produced by $\Phi$ decay will mainly form     $B_d$, $B^\pm$ and $B_s$ mesons. The fragmentation ratio of $b$-quarks to $B_d^0$, $B^\pm$ and $B_s^0$ is taken to be   4:4:1, which is roughly consistent with observation in $Z$ decays and $p-\bar{p}$ collisions \cite{PhysRevD.98.030001} (Note that the ratio is production-mode dependent with a slightly higher ratio of $B_s$ mesons produced in $p-\bar{p}$).  The charged $B$ mesons play no  role in baryogenesis, while the neutral $B$ mesons, as described in  Sec. \ref{s1},   will undergo  CP and flavor changing oscillations, and also sometimes decay into baryons and anti-baryons, as shown in FIG. \ref{decay}.

Since $B_d^0$ and $\overline{B}{}_d^0$ oscillate coherently while also potentially undergoing decoherence from scattering,    a density matrix treatment is useful for treatment of these states in the corresponding Boltzmann equations. Accounting for the interaction with plasma and the annihilation between $B_d^0$ and $\overline{B}{}_d^0$, the Boltzmann equations are \cite{Tulin:2012re,Cirelli:2011ac,Ipek:2016bpf}:
\begin{equation}
\begin{split}
\frac{dn}{dt}+3Hn=&-i(\mathcal{H} n-n\mathcal{H}^{\dagger} )-\frac{\Gamma_{\pm}}{2}[O_{\pm},[O_{\pm},n]]  \\
& -\langle\sigma v\rangle_{\pm}\left(\frac{1}{2}\{n, O_{\pm}\bar{n}O_{\pm}\}-n_{\text{eq}}^2\right) \\
&+\frac{1}{2}\frac{\Gamma_{\Phi}\rho_{\Phi}}{m_{\Phi}}\text{Br}_{\Phi \rightarrow B} O_{+}
 \end{split}
\end{equation}
 where the last term describes $B_d^0$ and $\overline{B}{}_d^0$ production during the decay of $\Phi$. 
 
 Here $\text{Br}_{\Phi \rightarrow B} $ is the branching ratio for $\Phi \rightarrow B+X$, which is assumed to be one because $\Phi$ mainly decays to b quarks and b quarks mainly hadronize to light B mesons $B_d^0$ and $B_s^0$. This is actually the very reason that we need $B$ meson oscillations to explain baryogenesis from the theoretical point of view. In this equation $n$ and $\bar{n}$ are density matrices,
 \begin{equation}
 n=\begin{pmatrix}
 n_{BB}&n_{B\bar{B}}\\
 n_{\bar{B}B}&n_{\bar{B}\bar{B}}
 \end{pmatrix}, \quad
  \bar{n}=\begin{pmatrix}
 n_{\bar{B}\bar{B}}&n_{B\bar{B}}\\
 n_{\bar{B}B}&n_{BB}
 \end{pmatrix},
 \end{equation}
and $n_{\text{eq}}$ is the equilibrium density of B mesons plus anti B mesons.  $\mathcal{H}$ is the Hamiltonian  for $B-\overline{B}$ mixing, see in Eq.(\ref{Ha}).$ \langle\sigma v\rangle_{\pm}$ are thermally-averaged annihilation cross section for B meson and anti B mesons,  and $\Gamma_{\pm}$ are the scattering rates between B mesons and charged particles in the plasma. It turns out the annihilation is negligible for B mesons, while the scattering would be important because there is a charge radius for B mesons which is different for mesons and anti mesons and  interacts   with the $e^\pm$ particles in the plasma. $O_{\pm}$ is a matrix
\begin{equation}
O_{\pm}=\begin{pmatrix}
1&0\\0& \pm1
\end{pmatrix}.
\end{equation}
The subscript $\pm$ is determined by the behavior of effective Lagrangian that gives rise to the interaction under charge conjugation of B mesons, $B\leftrightarrow\bar{B}$, $\mathcal{L}_{\text{eff}}\leftrightarrow \pm \mathcal{L}_{\text{eff}}$. Interactions that do not change sign are called flavor-blind interactions while those that change are flavor-sensitive interactions. For the concern of this work, the only interaction that is important to us  is the charge radius which gives the neutral $B$ mesons a photon coupling and allows for scattering off of $e^\pm$ particles.
 
\begin{equation}
\begin{split}
&\Sigma \equiv n_{BB}+n_{\bar{B}\bar{B}},\quad \Delta \equiv n_{BB}-n_{\bar{B}\bar{B}}, \\
& \Xi \equiv n_{B\bar{B}}-n_{\bar{B}B}, \quad  \Pi \equiv n_{B\bar{B}}+n_{\bar{B}B}.
\end{split}
\end{equation}
After hadronization, b quarks form $B_d^0$ , $B_s^0$ and other B mesons, with fragmentation ratio $\text{Br}(\bar{b}\rightarrow B_d^0)=0.4$ and  $\text{Br}(\bar{b}\rightarrow B_s^0)=0.1$\cite{PhysRevD.98.030001}.
Considering the flavor-sensitive interaction only, we can write the Boltzmann equations as:
\begin{equation}
\begin{split}
\left( \frac{d}{dt}+3H\right)\Sigma=&\frac{\Gamma_{\Phi}\rho_{\Phi}}{m_{\Phi}}\text{Br}_{\Phi \rightarrow B} -\Gamma_B\Sigma -(\text{Re}\Gamma_{12})\Pi \\ &+i(\text{Im}\Gamma_{12})\Xi, \\
\left( \frac{d}{dt}+3H\right)\Delta=&-\Gamma_B \Delta +2i(\text{Re}M_{12})\Xi +2(\text{Im}M_{12})\Pi,\\
\left( \frac{d}{dt}+3H\right)\Xi=&-(\Gamma_B+2\Gamma_{\text{sc}}) \Xi  
+2i(\text{Re}M_{12})\Delta\\ 
&-i(\text{Im}\Gamma_{12})\Sigma,\\
\left( \frac{d}{dt}+3H\right)\Pi =&-(\Gamma_B+2\Gamma_{\text{sc}})\Pi-2(\text{Im}M_{12})\Delta -(\text{Re}\Gamma_{12})\Sigma,
\end{split}
\end{equation}
where $\Gamma_B$ is the decay rate of B mesons, $M_{12},\Gamma_{12}$ are the off-diagonal terms of Hamiltonian described in Eq.(\ref{Ha}), and $\Gamma_{\Phi}$ is the decay rate of $\Phi$, which can define the reheaing temperature:
\begin{equation}
\Gamma_{\Phi}=3 H(T_{rh}),
\end{equation}
where the universe is assumed to be dominated by radiation with temperature $T_{rh}$. 

From these equations we can see coherent oscillations will cause transition $\Pi\rightarrow \Delta$, which will produce B meson asymmetry and thus baryon asymmetry as long as B mesons will decay to baryons. Flavor-sensitive scattering will suppress this process.

The scattering rate could be estimated as\cite{Elor:2018twp}:
\begin{equation}
\Gamma_{\text{sc}}\sim 10^{-11}\left( \frac{T}{0.02\text{GeV}} \right)^5\text{GeV}
\end{equation}

When T is below 0.01 GeV, the scattering rate is small and the decoherence caused by scattering is no longer significant. The production of the baryon asymmetry mainly takes place below this temperature.  

The baryon asymmetry is directly determined by B meson asymmetry:
\begin{equation}
\left( \frac{d}{dt}+3H\right)\delta_{\mathcal{B}}=\text{Br}_{B\rightarrow\mathcal{B}} \Gamma_B \Delta ,
\end{equation}
where $\delta_{\mathcal{B}}$ is the number density of baryon asymmetry and $\text{Br}_{B\rightarrow\mathcal{B}}$ is the branching ratio for $B\rightarrow\mathcal{B}+X$.

It is worth noting that the lifetime and oscillation period of B mesons are much shorter than the Hubble time, so the Hubble term and the time derivative term could be ignored. Under this approximation, we can compute the ratio $\Delta/\rho_{\Phi}$, which is a function of transition amplitudes and scattering rate:
\begin{equation}
\begin{split}
\frac{\Delta}{\rho_{\Phi}}&=
2|M_{12}||\Gamma_{12}|\text{sin}(\phi_{\Gamma}-\phi_M)\Gamma_{\Phi}(\Gamma_B+2\Gamma_{\text{sc}})\\
&\times\{-4|M_{12}|^2|\Gamma_{12}|^2\text{cos}(\phi_{\Gamma}-\phi_M)\\
&+(\Gamma_B^2+2\Gamma_B\Gamma_{\text{sc}})(4|M_{12}|^2-|\Gamma_{12}|^2)\\
&+\Gamma_B^2(\Gamma_B+2\Gamma_{\text{sc}})^2\}^{-1},
\end{split}
\end{equation}

where $\phi_{\Gamma}$, $\phi_M$ is the phase of $\Gamma_{12}$ and $M_{12}$. Under the approximation of $\text{cos}(\phi_{\Gamma}-\phi_M)\approx1$,  $\Delta/\rho_{\Phi}$ is proportional to $A_{\text{sl}}$. Given reheating temperature of the late decaying particle $\Phi$, the number density of net baryons could be calculated by: 
\begin{equation}
n_{\mathcal{B}}=\int_{t_0}^{t_1}\text{Br}_{B\rightarrow\mathcal{B}}\frac{\Delta}{\rho_{\Phi}}(t)\frac{\rho_{\Phi}(t)}{m_{\Phi}}\Gamma_B\frac{R(t)^3}{R(t_1)^3} dt,
\end{equation}
where $R(t)$ is the scale factor of the universe. $\Delta/\rho_{\Phi}$ is a function of transition amplitudes and scattering rate, and only scattering rate depends on temperature, which is a function of time. Therefore  $\Delta/\rho_{\Phi}$ itself is also a function of time. Here $t_0$ is the time when hadronization begins and $t_1$ is some time when the universe is dominated by radiation again. The baryon asymmetry would be given by:
\begin{equation}
Y=\frac{ n_{\mathcal{B}}}{\frac{2\pi^2}{45} g_{*s}(t_1)T(t_1)^3},
\end{equation}
where $g_{*s}$ is the effective degree of freedom at present. 
The baryon asymmetry can be represented as:
\begin{equation}
Y=\left(\frac{\text{Br}_{B\rightarrow\mathcal{B}}}{10^{-2}}\right)\left(\frac{100\text{GeV}}{m_{\Phi}}\right)(\alpha(T)A_{\text{SL}}^d+\beta(T)A_{\text{SL}}^s),
\end{equation}
where $\alpha(T)$ and $\beta(T)$ are coefficients as a function of reheating temperature $T$, whose exact values request numerical study. 
  \begin{figure}[h] 
 \includegraphics[width=\linewidth]{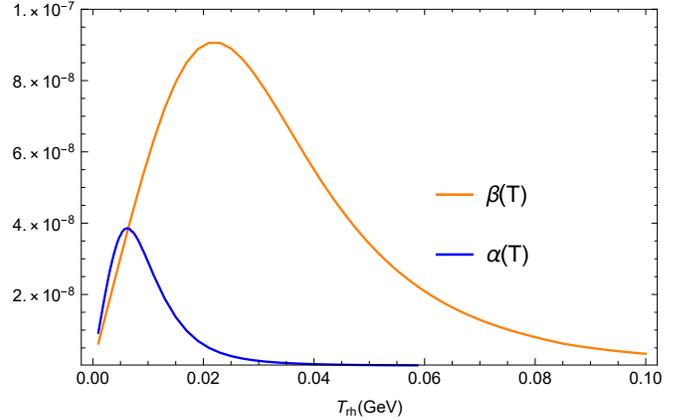}
 \caption{This figure shows the ability of $B_d^0$ and $B_s^0$ to produce baryon number at a given reheating temperature. 
 The scattering rate is greater at higher temperature, which leads to more significant decoherence and suppression of the baryon asymmetry. However,  $|M_{12}|$ and $|\Gamma_{12}|$ are very different for $B_d^0$ and $B_s^0$ , which leads to a lower characteristic temperature when decoherence is significant for the $B_d$ than the $B_s$.} Thus typically the baryon asymmetry is mostly produced in $B_s$ oscillations.
 \label{baryon}
 \end{figure}
 
The baryon asymmetry, with measurement from Cosmic Microwave
Background (CMB)\cite{Ade:2015xua,Aghanim:2018eyx} and Big Bang Nucleosynthesis (BBN) \cite{Cyburt:2015mya,Tanabashi:2018oca}, is given by :
\begin{equation}
Y= (8.718\pm 0.004)\times10^{-11}
\end{equation}
From FIG.(\ref{baryon}) we can conclude that SM predictions of $A_{\text{SL}}$ is too small to produce the expected baryon number. Besides,the SM CP violation predicts that the $B_d$ makes a negative contribution to baryon number. The fact that the $B_d$ contribution is suppressed down to lower temperature means one might hope to produce the correct sign from $B_s$ oscillations, however detailed computation  shows that   the net effect from the SM CP violation is to   give the wrong sign for the baryon asymmetry.  

We require $A_{\text{SL}}$ to be greater than the SM prediction and  positive for the $B_s$, and not too negative for the $B_d$, which is a testable feature of our mechanism. However, the exact constraints sensitively depend  on the reheating temperature, as shown in FIG. (\ref{baryon}). When reheating temperature is at 5 MeV, the $A_{\text{SL}}$ should satisfy:
\begin{equation}
\left(\frac{\text{Br}_{B\rightarrow\mathcal{B}}}{10^{-2}}\right)\left(\frac{100\text{GeV}}{m_{\Phi}}\right)(0.42A_{\text{SL}}^d+0.35A_{\text{SL}}^s)\approx 10^{-3}.
\end{equation}
When reheating temperature is higher, the contribution from $B_d$ would be suppressed. For reheating temperature at 25 MeV, the $A_{\text{SL}}$ should satisfy: 
\begin{equation}
\left(\frac{\text{Br}_{B\rightarrow\mathcal{B}}}{10^{-2}}\right)\left(\frac{100\text{GeV}}{m_{\Phi}}\right)(0.03A_{\text{SL}}^d+1.02A_{\text{SL}}^s)\approx  10^{-3}.
\end{equation}
 Given that $A_{\text{SL}}$ is always negative for the $B_d$, a high reheating temperature around 25 MeV is needed in order to suppress the  CP violation in the $B_d$ oscilaltions. Combining with the range of allowed $A_{\text{SL}}$ in Eq.(\ref{ASL}), we find that a branching ratio $\text{Br}_{B\rightarrow\mathcal{B}}$ larger than about $10^{-3}$ is required for baryon violating $b$ quark decays. Also,  To produce the expected baryon number, we expect $A_{\text{SL}}^s$ to be positive and greater than the SM predictions.

\section{Discussion}\label{s3}

In this work, we've shown that B meson oscillations could solve the puzzle of baryogenesis within a simple renormalizable model containing three new particles. This model predicts an exotic baryon number violating  B meson  decay mode. We also predict new contributions  to the semileptonic asymmetry in $B$ meson oscillations. Specifically, we  link the sign of the matter asymmetry to a new positive contribution to the semileptonic symmetry in  $B_s$ meson decays.  Another prediction is that immediately prior to nucleosynthesis in the early universe, the energy density is dominated by a massive late decaying particle. This may have implications for forming small clumps of dark matter  in the early universe.  

The baryon asymmetry is proportional to the branching fraction for  exotic B meson decays into a baryon and missing energy. These features could be searched at LHCb and Belle-\rom{2}, as discussed in \cite{Elor:2018twp}. The missing energy is carried by a long-lived Majorana fermion  $\chi$, which will  decay into a baryon or anti-baryon, and could be found in dedicated searches for long-lived particles \cite{Curtin:2018mvb,Lee:2018pag}. 
  
 The cosmological constraints for the corresponding reheating temperature is $T_{\text{RH}}>4.7\text{MeV}$\cite{deSalas:2015glj}. The thermal history of early universe before BBN is hard to probe. A late decaying $\phi$ particle does have some implications for axion miniclusters\cite{Nelson:2018via} and substructure formation\cite{Erickcek:2011us}. It happens that the reheating temperature that is favored by axion minicluster also tends to generate baryon number efficiently with our mechanism. Looking for axion miniclusters may also provide evidence for early matter domination.  
 
 Embedding this model into a R-parity violating supersymmetric model would provide  more motivation  and phenomenological implications.

\section*{Acknowledgements}
This work was supported in part by the U.S. Department of Energy, under grant number  DE-SC0011637. 
AEN is also supported in part by the Kenneth K. Young Memorial Endowed Chair. AEN acknowledges the hospitality of the Aspen Center for Physics, which is supported by National Science Foundation grant PHY-1607611.

\appendix*
\section{A}
In this appendix we will give explicit formulae for the extra contributions to $M_{12}$ from new physics, whose Feynman diagrams are already shown in Fig. (\ref{box}). They are very similar diagrams with those from Standard Model except the gamma matrices. Therefore we are   doing similar computations as in refs. \cite{Cheng:1982hq,Hagelin:1981zk,Buras:1984pq}.

The first box diagram will give us the following effective Hamiltonian:
\begin{equation}
H_{\text{eff}}^{(1)}= \overline{b}{}_v(1-\gamma^5)\gamma^{\alpha}d_u\overline{d}{}_v(1-\gamma^5)\gamma^{\beta}b_u p_{\alpha}(p-q)_{\beta}I_{\alpha\beta}^{(1)}+\text{h.c.}
\end{equation}
where $p$ is the momentum of b quark and $q$ is the loop momentum. The integral $I_{\alpha\beta}$ is given as:

\begin{equation}
\begin{split}
I_{\alpha\beta}^{(1)}=\frac{1}{4}(y_b y_d^{*})^2 &\int \frac{d^4q}{(2\pi)^4} q_{\alpha}(p-q)_{\beta}
\frac{1}{(q^2+m_{\phi}^2)}\\
&\times\frac{1}{[(p-q)^2+m_{\phi}^2]}\\
&\times\frac{1}{(q^2+m^2)[(p-q)^2+m^2]},
\end{split}
\end{equation}
where $m$ denotes the mass of Majorana fermion. We've ignored the momentum distributed to u,s quarks because they can be approximated as massless particles.
The second Feynman diagram will give us similar results. However, the contribution from the third diagram is negligible. After manipulating those gamma matrices, we will find that it is proportional to $m^2$, which is very small compared to $m_{\phi}^2$. The effective Hamiltonian from the second diagram could be written as:
\begin{equation}
H_{\text{eff}}^{(2)}=\overline{b}{}_{v}(1-\gamma^5)\gamma^{\alpha}d_{v}\overline{b}{}_{u}(1-\gamma^5)\gamma^{\beta}d_u(p-q)_{\alpha}q_{\beta} I^{(2)}_{\alpha\beta}+\text{h.c.}
\end{equation}
Switching indices on $I_{\alpha\beta}^{(1)}$ will give us $I_{\alpha\beta}^{(2)}$:
\begin{equation}
I_{\alpha\beta}^{(2)}=I_{\beta\alpha}^{(1)}
\end{equation}
There is a trick on calculating this Feynman integral, which utilize the symmetry of the Feynman diagrams.
\begin{equation}
\begin{split}
I_{\alpha\beta}^{(1)}=&q_{\alpha}(p-q)_{\beta}\frac{1}{(m_{\phi}^2-m^2)}\int \frac{d^4q}{(2\pi)^4}\\
&\times \{\frac{1}{(q^2+m^2)[(p-q)^2+m^2]} \\
&+\frac{1}{(q^2+m_{\phi}^2)[(p-q)^2+m_{\phi}^2]}\\
&- \frac{1}{(q^2+m_{\phi}^2)[(p-q)^2+m^2]}\\
&-\frac{1}{(q^2+m^2)[(p-q)^2+m_{\phi}^2]}
\}
\end{split}
\end{equation}
Applying Feynman's formula:
\begin{equation}
\begin{split}
&\frac{1}{(q^2+m_1^2)[(p-q)^2+m_{2}^2]}=\\
&\int_0^1 dx 
 \{  (q-xp)^2+x(1-x)p^2+xm_2^2+(1-x)m_1^2   \}^{-2}.
\end{split}
\end{equation}
Since the momentum is given by on-shell relation: $p^2=-m_b^2$, it is convenient to denote:
\begin{equation}
\begin{split}
&D_1=-x(1-x)m_b^2+m_{\phi}^2\\
&D_2=-x(1-x)m_b^2+m^2\\
&D_3=-x(1-x)m_b^2+xm^2+(1-x)m_{\phi}^2\\
&D_4=-x(1-x)m_b^2+xm_{\phi}^2+(1-x)m^2
\end{split}
\end{equation}
The Hamiltonian could be represented as:
\begin{equation}
\begin{split}
\mathcal{H}_{\text{eff}}=&A\overline{d}{}_{\alpha} \gamma^{\mu}(1+\gamma^5) b_{\alpha}\overline{d}{}_{\beta} \gamma_{\mu}(1+\gamma^5)b_{\beta} +\\
&+B\overline{d}{}_{\alpha}(1-\gamma^5)b_{\alpha}\overline{d}{}_{\beta}(1-\gamma^5)d_{\beta}+\text{h.c.}
\end{split}
\end{equation}
Coefficients $A, B$ can be written as:
\begin{equation}
\begin{split}
&A=\frac{-1}{16\pi^2(m_{\phi}^2-m^2)^2}\int_0^1dx{\sum_{i=1}^4}^{\prime} D_i \text{ln} \frac{D_i}{m_{\phi}^2} \\
&B=\frac{m_b^2}{32\pi^2(m_{\phi}^2-m^2)^2}\int_0^1 dx{\sum_{i=1}^4}^{\prime}\text{ln}D_i ,
\end{split}
\end{equation}
where  $
{\sum_{i=1}^4}^{\prime}={\sum_{i=1}^2}-{\sum_{i=3}^4}
$.
Since $D_i$ are always positive for the whole range of $x$, there is no contribution to $\Gamma_{12}$. The contribution to $M_{12}$ is:
\begin{equation}
\Delta M_{12}=\langle B^0| \mathcal{H}_{\text{eff}}|\overline{B}{}^0\rangle.
\end{equation}
The matrix elements are corresponding to the nonperturbative effects, which could be estimated as:
\begin{equation}
\begin{split}
&\langle B^0|\overline{d}{}_{\alpha} \gamma^{\mu}(1+\gamma^5) b_{\alpha}\overline{d}{}_{\beta} \gamma_{\mu}(1+\gamma^5)b_{\beta}|\overline{B}{}^0\rangle=\frac{8}{6} f_B^2 m_B B_B^2\\
&\langle B^0|\overline{d}{}_{\alpha}(1-\gamma^5)b_{\alpha}\overline{d}{}_{\beta}(1-\gamma^5)d_{\beta}|\overline{B}{}^0\rangle =-\frac{5}{6} f_B^2 m_B B_B^2,
\end{split}
\end{equation}
 where $f_B$ is the decay constant of B mesons, $m_B$ is B meson mass and $B_B$ is a bag parameter, which is order one. 
 For $m=2\text{GeV}$, which is the parameter space we are interested in, the result is:

 \begin{equation}
\Delta M_{12}= 0.66\frac{f_B^2m_B B_B^2 }{16\pi^2 m_{\phi}^2}(y_b y_d^*)^2  
 \end{equation}
 Lattice study can numerically give us the value of decay constant and bag parameter, $B_B=0.87, B_{B_s}=0.9, f_B=0.192\text{GeV}, f_{B_s}=0.228\text{GeV}$. \cite{Carrasco:2012ps,Lucha:2013vwa}
 
 The ratio to the experimental value of $M_{12}$ is:
 \begin{equation}
 \begin{split}
&  \frac{\Delta M_{12}}{M_{12}}=3689\left(\frac{1\text{TeV}}{m_{\phi}}\right)^2(y_b y_d^*)^2 ,
\\
&\frac{\Delta M_{12s}}{M_{12s}}=161\left(\frac{1\text{TeV}}{m_{\phi}}\right)^2(y_b y_s^*)^2 .
 \end{split}
 \end{equation}
 
 Therefore $y_b, y_d, y_s$ must be much smaller than 1 to agree with experimental results. This will not cause a problem because $\Delta M_{12}$ is the only observable that is only determined by $y_d$. It is worth noting that the calculations we present here are very much the same with box diagram calculations in a supersymmetric theory\cite{Ciuchini:1998ix}, but with very different parameters.

\bibliographystyle{apsrev}
\bibliography {baryogenesis.bib}

\end{document}